\documentclass[sn-mathphys,Numbered]{sn-jnl}

\usepackage{graphicx}%
\usepackage{multirow}%
\usepackage{amsmath,amssymb,amsfonts}%
\usepackage{amsthm}%
\usepackage{mathrsfs}%
\usepackage[title]{appendix}%
\usepackage{xcolor}%
\usepackage{textcomp}%
\usepackage{manyfoot}%
\usepackage{booktabs}%
\usepackage{algpseudocode}%
\usepackage{listings}%
\usepackage{tikz}
\usepackage{float}
\usepackage{algorithm2e}
\usepackage{ragged2e}
\usepackage{natbib}

 \newtheorem{example}{Example}%
\newtheorem{definition}{Definition}%

\begin{document}

\title[On Graph Grammars and Games]{On Graph Grammars and Games}


\author[1,2]{\fnm{Jayakrishna} \sur{Vijayakumar}}\email{vjayakrishna@amaljyothi.ac.in}
\equalcont{These authors contributed equally to this work.}
\author*[3]{\fnm{Lisa} \sur{Mathew}}\email{lisamathew@amaljyothi.ac.in}
\equalcont{These authors contributed equally to this work.}

\affil[1]{\orgdiv{Department of Computer Science and Engineering}, \orgname{Amal Jyothi College of Engineering}, \orgaddress{\street{Kanjirappally}, \city{Kottayam}, \postcode{686 518}, \state{Kerala}, \country{India}}}

\affil[2]{\orgdiv{Research scholar}, \orgname{APJ Abdul Kalam Technological University}, \orgaddress{\street{CET Campus}, \city{Thiruvananthapuram}, \postcode{695 016}, \state{Kerala}, \country{India}}}

\affil[3]{\orgdiv{Department of Basic Sciences }, \orgname{Amal Jyothi College of Engineering}, \orgaddress{\street{Kanjirappally}, \city{Kottayam}, \postcode{686 518}, \state{Kerala}, \country{India}}}

\abstract{Graph grammars  form an interesting area of research because of their versatility in modelling diverse situations with graphs as the structures which are to be manipulated. A new class of graph grammars, \textit{nc-eNCE Graph Grammars} has been introduced recently with an aim of restricting the order of application  of graph production rules, thereby generating different graph classes using the same set of rules. On the other hand 2D game design using an algorithmic approach known as procedural content generation has been of interest recently. In this paper we modify the structure of nc-eNCE graph grammars with the aim of generating directed graphs. We show that employing these graph grammars simplifies the design of 2D games. We have also developed an algorithm which makes use of these graph grammars for generating random game level layouts ensuring that the players will get a different gaming experience each time they play.
}




\maketitle
3\section{Introduction}

Puzzle games form a fascinating  class of 2D games due to their challenging  and entertaining nature. However the creation of these games traditionally required immense manpower and time, thereby escalating the cost of developing these games. In order to avoid boredom on the part of players a game developer needs to provide new domains to the player so that the game becomes unpredictable. Making use of algorithms for creating these environments helps developers to produce new challenges at a lower cost. In this paper we look at certain categories of 2D games - specifically games having lock-and-key\cite{Ashmore_2007} patterns as challenges.  
\par Any system where items are interconnected in some way, such as a network of computers, a city and its roads\cite{parish2001procedural}, or a group of friends on a social network, aquatic predatory networks, plant pollinator networks, etc. can be represented using graphs. As the game design methodology used here is algorithmic in nature, the best data structure that complements this approach is a graph.  
 In this paper we employ graph grammars as a mechanism for creating these game plots/levels. We have designed an algorithm which randomly generates a new game plot each time the game is played, ensuring a stimulating experience for the players.
\par The rest of this paper is organised as follows. Section \ref{prelims} presents   the key  concepts associated with game design and graph grammars   required for understanding the rest of this paper. Sections \ref{Dr-ncence} and \ref{djump} develop a modified  version of the nc-eNCE and jumping graph grammars introduced in \cite{jk_icmicds22,jk_symetry} so as to deal with directed graphs. Finally Section \ref{game}  deals with two dimensional game generation and shows how the modified graph grammars discussed in  Sections \ref{Dr-ncence} and \ref{djump} can be used to generate solvable game plots. It also includes an algorithm for generating these plots randomly.
\section{Preliminaries}\label{prelims}
 
\subsection{2D Games}
We  present some of the fundamental terminology associated with game design.
\begin{itemize}
\item[] \textbf{Procedural content generation\cite{hendrikx2013} -} the process of using  algorithmic concepts to produce game content. 
\item[] \textbf{Level/Plot\cite{rouse2004} - } Specific area in a gaming environment restricted by the architecture of the game space.
\item[] \textbf{Linear Games\cite{adams2014} - } Games where a player travel over the gaming world in a well established order. The levels/ plots are kept in such a way that a player can only move to a unique plot from  his/her current space.
\item[] \textbf{Action-Adventure Games\cite{dormans2010}-} 2D games that have warfare, investigations and puzzle solving. It is typical to incorporate obstacles like mines and fights into a puzzle scenario while setting up an action-adventure game. Classic video games like \textit{Beyond Good and Evil\text{,} Constantine}, etc. come under this category. 
\item[] \textbf{Logic Puzzles -} Solvable challenges which make use of reasoning based only on some prior information which has been provided within the puzzle.
\item[] \textbf{Missions -} Circumstances in which a player will move along the game environment in order to achieve a specific target by surmounting a number of hurdles.
\item[] \textbf{Platform Games - } Games where barriers in a path can be avoided by jumping  between different platforms.
\end{itemize}
\subsection{Lock-and-Key Puzzles}
The games level  designs considered in this paper have some common elements and logic. They  can be referred to as lock-and-key puzzles\cite{dormans2011}. The components of  a lock-and-key puzzle are explained here.
\subsubsection{Structure}
The structure of these puzzles consists of objects in the form of locks and keys\cite{Ashmore_2007}. A lock is a barrier that restricts the player from advancing across a level unless a condition is satisfied. A lock has an orientation and direction. Once it has been opened, the player gains access to a distinct traversal. Some locks only work in one direction and prevent players from returning to the previous traversal plot they entered from once they have passed through. In order to access the lock the player should locate the key. A key is an item used  to open a lock. They can be in the form of switches or objects. Switches are keys that can open locks without moving them. Keys in the form of objects need to be selected and placed in the specific area to open a specific lock. If a player needs to complete a level or goal, then he needs to open one or more locks which prevents him from achieving it. A player cannot open a lock unless  he has the unique key for that lock and this becomes the central idea of the puzzle. A lock and key in the actual game world can be any pair of objects that satisfies the above said properties. A switch to open a door, a  trigger to blast open a cave, a vessel to move across a swift river, etc. are some scenarios which satisfies the lock and key logic. If it appears that there is no major obstruction to be considered in the way, a mere lock and key may be present in the space and be regarded as a part of the traversal. Even though locks and keys appear in a variety of ways in the environment of a game, their relationship does not change. The objects that are to be placed in a puzzle environment layout are of great importance as far as game playing is concerned.
\subsubsection{Traversal}
A traversal is considered as an interconnected area where a player has the freedom to move his character to and from any place in that space utilizing the in-game tools provided. There could be many rooms in that real game space scenario. On entering a traversal space\cite{dormans2010}, the player is presented with the barriers, challenges and constraints that makes the movement a challenging task. The space  contains difficult situations with obstacles in the form of fighting and platforms for agile movements. A player need to acquire a better understanding of the plot and a competence to overcome these situations. In this paper the focus is on abstracting the physical design and concentrating more on the structure of the game level. We are only interested in the puzzle's crucial path for the sake of this study. This implies that the player may ignore any portion of the mission that they are not needed to complete.
\subsubsection{Puzzle Graphs}
We use graph theory to present a conceptual description of game level and its structure . This is  referred to as  a puzzle graph environment. It is a directed graph with labelled nodes of any of the following four styles.
\begin{itemize}
    \item[] \textbf{Begin - } At this node the player begins the game/level. Movement to the next traversal point happens after that.
    \item[] \textbf{End - } The puzzle finishes if the player arrives at this node.
    \item[] \textbf{Lock - } A player's movement flexibility is restricted by these nodes. A lock node can be passed through only if all the respective key nodes  are activated. 
    \item[] \textbf{Key - } When there is an edge  traversing  which leads to this node, then a player can use the key node and use it for unlocking the corresponding lock node. Usually key nodes are kept in an active mode in most  game designs. 
\end{itemize}
We aim to design solvable puzzle graphs which ensures that the game terminates in a finite period of time. This is confirmed  by checking 
 whether there is a directed path from the begin node to the end node. An unsolvable puzzle scenario arises when  a key  for opening a lock  is kept behind the lock node and  hence is unreachable. There can also be situations where a lock does not have a key at all, preventing the forward movement of the player  thereby  making the end node unreachable. We  use an algorithmic approach towards  generation of solvable puzzle graphs.
\subsection{Graph Grammars}
The concept of a grammars in formal language theory \cite{hopcroft1979introduction} has motivated exploration  into the possibilities of their application to  more complex structures. Research has proved that this experimentation is worthwhile  for instance by utilizing  grammar theory  in the generation of planned blueprints of cities and buildings. Similarly the L- Systems \cite{linden} proposed by Lindenmayer has the ability to create structures  that mimic the natural growth of plants. 
\par Graph grammars came about as a result of a graph based approach to formal grammar theory \cite{rozenberg1997handbook}. Unlike a normal grammar which generates a collection of strings, a graph grammar is capable of generating a family of graphs which exhibit some specific characteristics. Graph grammars basically come under two categories namely edge replacement and node replacement graph grammars \cite{Rozenberg_97}. Edge/ Hyperedge replacement graph grammars \cite{Kreowski1986} typically are used to build hypergraphs wherein  multiple nodes get connected to an edge. \par There are three parts to a production rule for a graph grammar: the mother and daughter subgraphs, and the embedding technique \cite{Rozenberg_97}. The host graph is the one to which we apply a production. If the mother subgraph is present in the host graph, a production can be applied to that graph. The daughter subgraph is substituted for the the mother subgraph in the host graph. The remainder of the host graph is then connected to the daughter subgraph in accordance with the embedding procedure. Formally a graph grammar production rule is of the form $(M,D,E)$ \cite{Rozenberg_97}. In this construct $M$ is the mother subgraph that is present in the host graph $H$ to which we apply the graph production rule. $D$ is the subgraph that replaces $M$ in $H$. $E$ is the embedding procedure that determines the new connections between $D$ and the rest of $H$. 
\par Graph grammars primarily have two embedding methods- gluing and connecting, or in other terms algebraic and algorithmic approaches \cite{Ehrig_1978,Ehrig_1986}. More on these concepts can be found in \cite{rozenberg1997handbook}. Meanwhile a few new variants of graph grammars had been introduced in \cite{jk_symetry}. We enhance the concepts of those graph grammars in order to generate directed graphs which in turn become the base in setting up the plot for game levels.

As this work involves the usage of node replacement graph grammars \cite{Rozenberg_97} and its enhanced variants, the   mother graph consists of a single node. The following example illustrates  the subgraph embedding procedure. 
\begin{example}
Consider the host graph shown in Figure \ref{host} and the graph production rule in Figure \ref{prod}. A production rule $p$ is applied only if we have a  subgraph in the host graph isomorphic to the graph  on the left of the production rule. Application of a production rule works in two stages. First we need to remove the mother node and then we need to embed the right  side of the corresponding production rules using the connection instructions associated with that production rule. As a result of this we generate a new graph. In our example, applying production rule $p$ on the host graph removes the node $B$ from the host graph and embeds the subgraph to the right of $p$ in the remaining  portion of the host graph using the connection instruction  $C=\{(A,\alpha\mid\beta,3)\}$. This instruction establishes edges labelled $\beta$ between the node labelled $A$ which was previously connected to $B$ using edge $\alpha$ and each node numbered $3$in the daughter graph. The two stages are shown in Figure \ref{embed}.
    \begin{figure}[H]
    \centering
    \begin{tikzpicture}
      [node distance=1.5cm,main node/.style={minimum size = .5cm,circle,fill=black!80,anchor=west,draw}, scale=1, every node/.style={scale=.7}]
\node[main node] (A1) [label={$A$}]{\textcolor{white}{\small{1}}};
\node[main node] (B) [right of=A1,label={$B$}]{\textcolor{white}{\small{2}}};
\node[main node] (A2) [right of=B,label={$A$}]{\textcolor{white}{\small{1}}};
\draw [thick] (A1.east) -- (B.west)node [above,midway] {$\alpha$};
\draw [thick] (B.east) -- (A2.west)node [above,midway] {$\alpha$};
   \end{tikzpicture}
    \caption{Host graph}
    \label{host}
\end{figure}
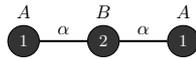
\vspace{-1.2cm}
  \begin{figure}[H]
    \centering
    \begin{tikzpicture}
      [node distance=1.5cm,main node/.style={minimum size = .5cm,circle,fill=black!80,anchor=west,draw}, scale=.8, every node/.style={scale=.7}]
\node (p1) {$p:$}{};
\node[main node] (C1) [right of= p1,label={$B$}]{\textcolor{white}{\small{2}}};
\node (d1) [right of= C1]{$\rightarrow$};
\node[main node] (a) [right  of=d1,label=left:{$C$}]{\textcolor{white}{\small{3}}}; 
\node[main node] (rC) [below  of=a,label=below:{$B$}]{\textcolor{white}{\small{2}}}; 
 \node (C) [right of=a,anchor=west]{$C=\{(A,\alpha\mid\beta,3)\}$};
 \node[main node] (uB) [above  of=a,label={$B$}]{\textcolor{white}{\small{2}}}; 
 \draw [thick] (a) -- (rC) node [right,midway] {$\alpha$};
  \draw [thick] (a) -- (uB) node [right,midway] {$\alpha$};
   \end{tikzpicture}
    \caption{Graph Production Rule}
    \label{prod}
\end{figure}
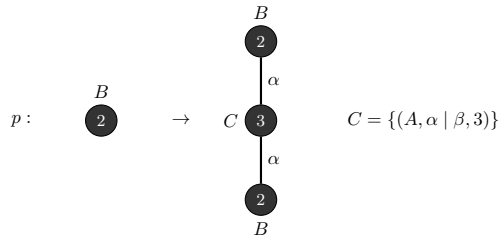
\vspace{-1cm}
    \begin{figure}[H]
    \centering
    \begin{tikzpicture}
      [node distance=1.5cm,main node/.style={minimum size = .5cm,circle,fill=black!80,anchor=west,draw}, scale=0.8, every node/.style={scale=.7}]
\node[main node] (A1) [label={$A$}]{\textcolor{white}{\small{1}}};
\node(D)[right of =A1]{};
\node[main node] (A2) [right of=D,label={$A$}]{\textcolor{white}{\small{1}}};
\node(D)[right of =A2]{};
\node(D)[right of =D]{};
\node[main node] (A11) [right of=D,label=left:{$A$}]{\textcolor{white}{\small{1}}};
\node[main node] (C) [right of=A11,label=above right:{$C$}]{\textcolor{white}{\small{3}}};
\node[main node] (A12) [right of=C,label=right:{$A$}]{\textcolor{white}{\small{1}}};
\node[main node] (B11) [above of=C,label={$B$}]{\textcolor{white}{\small{2}}};
\node[main node] (B12) [below of=C,label=below:{$B$}]{\textcolor{white}{\small{2}}};
\draw [thick] (A11.east) -- (C.west)node [below,midway] {$\beta$};
\draw [thick] (C.east) -- (A12.west)node [below,midway] {$\beta$};
\draw [thick] (B11.south) -- (C.north)node [left,midway] {$\alpha$};
\draw [thick] (C.south) -- (B12.north)node [left,midway] {$\alpha$};

   \end{tikzpicture}
    \caption{Graph after the removal of mother node  and after the embedding}
    \label{embed}
\end{figure}
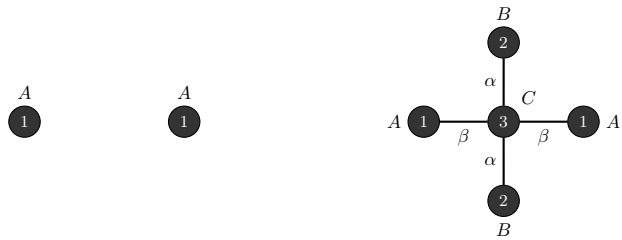
\end{example}
\par There are several variants of node replacement graph grammars \cite{Rozenberg_97,kukluk2007learning}. Most of these graph grammars are confluent. This means that there is no restriction in the order of application of production rules. In contrast, the new variants introduced in \cite{jk_symetry} has a function in the form of a regular expression $R(P)$ which controls the order in which the production rules  need to be applied in the host graph and thereby achieves nonconfluentiality in generation of undirected graph classes.

\section{Directed Non-Confluent Edge and Node Controlled Embedding (\textit{Dnc-eNCE}) Graph Grammar}
\label{Dr-ncence}
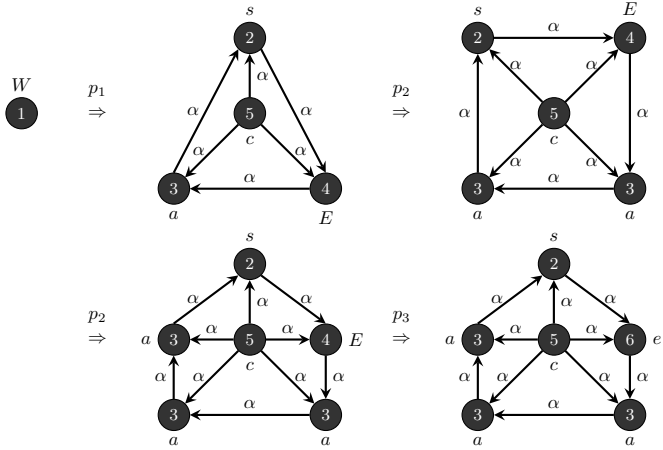
\begin{figure}[H]
  \begin{tikzpicture}[main node/.style={minimum size = .5cm,circle,fill=black!80 ,draw}, scale=0.3, every node/.style={scale=0.7,anchor=west}] 
 \matrix [column sep={1.cm,between origins},row sep={.5cm,between origins}]
 {
 
 &&&\node[main node](1)[label=$s$]{\textcolor{white} {\small{2}}};
 &&&\node[main node](6)[label=$s$]{\textcolor{white} {\small{2}}};\&&&
 \node[main node](7)[label=$E$]{\textcolor{white} {\small{4}}};
\\\\

\node[main node](2)[label=$W$]{\textcolor{white} {\small{1}}};&
\node [label=above:$p_1$]{$\Rightarrow$};&&
\node[main node](3)[label=below:$c$]{\textcolor{white} {\small{5}}};&&
\node [label=above:$p_2$] {$\Rightarrow$};&&
\node[main node](8)[label=below:$c$]{\textcolor{white} {\small{5}}};
\\\\
&&\node[main node](4)[label=below:$a$]{\textcolor{white} {\small{3}}};&&
\node[main node](5)[label=below:$E$]{\textcolor{white} {\small{4}}};&&
\node[main node](9)[label=below:$a$]{\textcolor{white} {\small{3}}};&&
\node[main node](10)[label=below:$a$]{\textcolor{white} {\small{3}}};\\\\
&&&\node[main node](11)[label=$s$]{\textcolor{white} {\small{2}}};&&&&
\node[main node](17)[label=$s$]{\textcolor{white} {\small{2}}};\\\\
& \node [label=above:$p_2$] {$\Rightarrow$};
&\node[main node](12)[label=left:$a$]{\textcolor{white} {\small{3}}};&
\node[main node](13)[label=below:$c$]{\textcolor{white} {\small{5}}};&
\node[main node](14)[label=right:$E$]{\textcolor{white} {\small{4}}};& 
\node [label=above:$p_3$] {$\Rightarrow$};&
\node[main node](18)[label=left:$a$]{\textcolor{white} {\small{3}}};&
\node[main node](19)[label=below:$c$]{\textcolor{white} {\small{5}}};&
\node[main node](20)[label=right:$e$]{\textcolor{white} {\small{6}}};
\\\\
&&\node[main node](15)[label=below:$a$]{\textcolor{white} {\small{3}}};& 
&\node[main node](16)[label=below:$a$]{\textcolor{white} {\small{3}}};&&
\node[main node](21)[label=below:$a$]{\textcolor{white} {\small{3}}};& 
&\node[main node](22)[label=below:$a$]{\textcolor{white} {\small{3}}};&\\};

\draw [thick,stealth-] (1.south west) -- (4.north) node [left,midway] {$\alpha$};
\draw [thick,-stealth] (1.south east) -- (5.north) node [right,midway] {$\alpha$};
\draw [thick,stealth-] (1.south) -- (3.north) node [right,midway] {$\alpha$};
\draw [thick,stealth-] (4.east) -- (5.west) node [above,midway] {$\alpha$};
\draw [thick,-stealth] (3.south west) -- (4.north east) node [left,midway] {$\alpha$};
\draw [thick,-stealth] (3.south east) -- (5.north west) node [right,midway] {$\alpha$};
\draw [thick,stealth-] (6.south) -- (9.north) node [left,midway] {$\alpha$};
\draw [thick,-stealth] (7.south) -- (10.north) node [right,midway] {$\alpha$};
\draw [thick,-stealth] (6.east) -- (7.west) node [above,midway] {$\alpha$};
\draw [thick,stealth-] (6.south east) -- (8.north west) node [above,midway] {$\alpha$};
\draw [thick,stealth-] (7.south west) -- (8.north east) node [above,midway] {$\alpha$};
\draw [thick,-stealth] (8.south west) -- (9.north east) node [below,midway] {$\alpha$};
\draw [thick,-stealth] (8.south east) -- (10.north west) node [below,midway] {$\alpha$};
\draw [thick,stealth-] (9.east) -- (10.west) node [above,midway] {$\alpha$};
\draw [thick,stealth-] (11.south west) -- (12.north) node [left,midway] {$\alpha$};
\draw [thick,-stealth] (11.south east) -- (14.north) node [right,midway] {$\alpha$};
\draw [thick,stealth-] (11.south) -- (13.north) node [right,midway] {$\alpha$};
\draw [thick,-stealth] (13.south west) -- (15.north east) node [left,midway] {$\alpha$};
\draw [thick,-stealth] (13.south east) -- (16.north west) node [right,midway] {$\alpha$};
\draw [thick,stealth-] (12.east) -- (13.west) node [above,midway] {$\alpha$};
\draw [thick,-stealth] (13.east) -- (14.west) node [above,midway] {$\alpha$};
\draw [thick,stealth-] (15.east) -- (16.west) node [above,midway] {$\alpha$};
\draw [thick,stealth-] (12.south) -- (15.north) node [left,midway] {$\alpha$};
\draw [thick,-stealth] (14.south) -- (16.north) node [right,midway] {$\alpha$};
\draw [thick,stealth-] (17.south west) -- (18.north) node [left,midway] {$\alpha$};
\draw [thick,-stealth] (17.south east) -- (20.north) node [right,midway] {$\alpha$};
\draw [thick,stealth-] (17.south) -- (19.north) node [right,midway] {$\alpha$};
\draw [thick,-stealth] (19.south west) -- (21.north east) node [left,midway] {$\alpha$};
\draw [thick,-stealth] (19.south east) -- (22.north west) node [right,midway] {$\alpha$};
\draw [thick,stealth-] (18.east) -- (19.west) node [above,midway] {$\alpha$};
\draw [thick,-stealth] (19.east) -- (20.west) node [above,midway] {$\alpha$};
\draw [thick,stealth-] (21.east) -- (22.west) node [above,midway] {$\alpha$};
\draw [thick,stealth-] (18.south) -- (21.north) node [left,midway] {$\alpha$};
\draw [thick,-stealth] (20.south) -- (22.north) node [right,midway] {$\alpha$};
\end{tikzpicture}
 \caption{Generation of the directed Wheel graph $W_6$.}
 \label{ddwheel}
 \end{figure}

In this section we present an enhanced variant of $nc-eNCE$ graph grammar introduced in \cite{jk_symetry} to derive the class of directed graphs. Here  an extra parameter is added to the connection instruction in  order to specify the direction of the newly created edges. We later use this new graph grammar variant to generate puzzle graphs \cite{Ashmore_2007} which becomes the base for 2D game design.
Formally we have the definition:
\begin{definition}\label{Dncence}
A construct $DrncG= ( \Sigma, \Delta, \Gamma, \Omega, P, G_S, R(P) )$ is known as an $Dnc-eNCE$ graph grammar where
\begin{itemize}
\item []$\Sigma$ and $\Gamma$ are sets of symbols used to label nodes and edges respectively,
\item []$\Delta $ and $\Omega$ are the collections of terminal symbols in $\Sigma$ and $\Gamma$ respectively, 
\item []A production rule in $P$, $p: A \rightarrow ( D, C )$ acting on the mother node $M$ with label $A$ has a collection C of connection instructions $(a,p\mid d\mid q ,B)$ associated with it. This indicates that the edge $p$ which connected $x$ (a neighbour of $M$ with label $a$) and $M$ is removed and a new edge $q$ is established between $x$ and $B$ (a node in $D$). The flag $d\in\{0,1,-1\}$ indicates the edge direction. $0$ indicates that the connection is bidirectional, $1$ indicate a directed edge from $x$ to $B$ and  $-1$ indicate a directed edge from $B$ to $x$.
\item []$G_S$ is the initial graph,
\item []The regular control, $R(P)$, regulates the sequence of application of the production rules.
\end{itemize}
\end{definition}
The graph grammar $ncG$ generates the language \[L(DrncG)={ \{ G \in G_\Delta |G_S \overset{R(P)}{\implies}G \} }\] 
Here, all the nodes of graphs in {$G_\Delta$} are labelled using $\Delta$ and $G_S \overset{*}{\Rightarrow} G $ is  as in definition 1. The ordered application of productions in the sequence $p_1, p_2, \cdots, p_n$ ($p_1p_2\cdots p_n \in L(R(P))$ leads to the generation of graphs specified by the language of the grammar.
\begin{example}\label{edwheel}
Consider $ncG_W= (\Sigma, \Delta, \Gamma, \Omega, P, G_S, R(P) )$, $\Sigma = \{ W,E,a,c,s,e\}$, $\Delta = \{ a,c,s,e\}$, $\Gamma = \{ \alpha \}$, $\Omega =\{\alpha\}$, $P = \{ p_1, p_2, p_3 \}$, $G_S$ is a single node labelled $W$ and R(P)= $p_1p_2^*p_3$.
The productions in \textit{P} 
 are shown in   Figure \ref{pdwheel}. 
 \label{exwheel} \end{example}
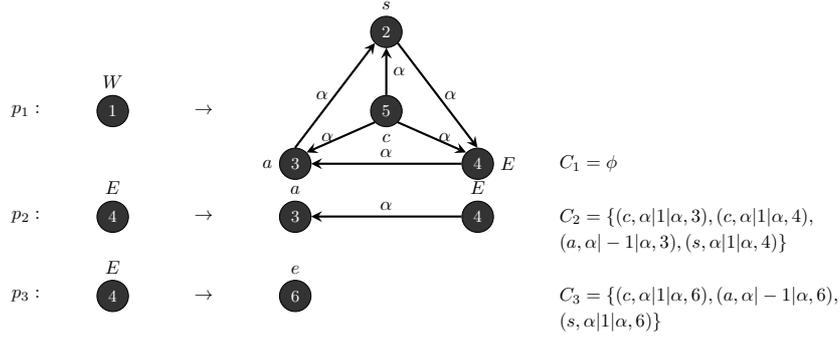
\begin{figure}[H]
 \begin{tikzpicture}[main node/.style={minimum size = .5cm,circle,fill=black!80 ,draw}, scale=0.3, every node/.style={scale=0.7,anchor=west}] 
 \matrix [column sep={1.2cm,between origins},row sep={.35cm,between origins}]
 {
 &&&&\node[main node](1)[label=$s$]{\textcolor{white} {\small{2}}};&
\\\\\\
\node {$p_1:$};&
\node[main node](2)[label=$W$]{\textcolor{white} {\small{1}}};&
\node{$\rightarrow$};&&
\node[main node](3)[label=below:$c$]{\textcolor{white} {\small{5}}};\\\\
&&&\node[main node](4)[label=left:$a$]{\textcolor{white} {\small{3}}};&&
\node[main node](5)[label=right:$E$]{\textcolor{white} {\small{4}}};&
\node {$C_1 = \phi$};\\\\
\node {$p_2:$};&
\node[main node](6)[label=$E$]{\textcolor{white} {\small{4}}};&
\node {$\rightarrow$};&
\node[main node](7)[label=$a$]{\textcolor{white} {\small{3}}};&&
\node[main node](8)[label=$E$]{\textcolor{white} {\small{4}}};&
\node{$C_2= \{(c,\alpha|1|\alpha,3),(c,\alpha|1|\alpha,4),$};\\
&&&&&&\node{$ (a,\alpha|-1|\alpha,3),(s,\alpha|1|\alpha,4)\}$};\\\\
\node {$p_3:$};&
\node[main node](9)[label=$E$]{\textcolor{white} {\small{4}}};&
\node {$\rightarrow$};&
\node[main node](10)[label=$e$]{\textcolor{white} {\small{6}}};&&&
\node{$C_3= \{(c,\alpha|1|\alpha,6),(a,\alpha|-1|\alpha,6),$};\\&&&&&&\node{$(s,\alpha|1|\alpha,6)\}$};\\};
\draw [thick,stealth-] (1.south west) -- (4.north) node [left,midway] {$\alpha$};
\draw [thick,-stealth] (1.south east) -- (5.north) node [right,midway] {$\alpha$};
\draw [thick,stealth-] (1.south) -- (3.north) node [right,midway] {$\alpha$};
\draw [thick,stealth-] (4.east) -- (5.west) node [above,midway] {$\alpha$};
\draw [thick,-stealth] (3.south west) -- (4.north east) node [left,midway] {$\alpha$};
\draw [thick,-stealth] (3.south east) -- (5.north west) node [right,midway] {$\alpha$};
\draw [thick,stealth-] (7.east) -- (8.west) node [above,midway] {$\alpha$};
\end{tikzpicture}
 \caption{Production rules for directed Wheel graph.}
 \label{pdwheel}
\end{figure}
Figure \ref{ddwheel} shows the generation of the directed Wheel graph $W_6$ using the grammar $ncG_W$.
\section{Directed Non-confluent Edge and Node Controlled Embedding Jumping Graph grammars (\textit{Dnc-eNCE-JGG})}\label{djump}
We extend the concept of $Drnc-eNCE$ graph grammar in Definition \ref{Dncence} as follows.
 Formally, we have:
\begin{definition} \label{Drjump}
A $Dnc-eNCE-JGG$ graph grammar is a 7 tuple: $DrncJGG= \text{(}\Sigma, \Delta, \Gamma, \Omega, $ $P, G_S, R\text{(}P\text{)} $\text{)} where
\begin{itemize}
\item []$\Sigma$ is the set of node labels,
\item []$\Delta$ is the set of terminal node labels,
\item []$\Gamma$ is the set of edge labels,
\item []$\Omega$ is the set of terminal  edge labels,
\item []The productions in $P$ are either as defined in \ref{Dncence} or of the  form $p:  A \rightarrow ( D, C )$ where $A$ is the label of the mother node and $D$ is the daughter graph.
The connection instruction $C$ can be in any one of the following forms
\begin{enumerate}
\item ($a,p\mid d\mid q ,B$), where $p$ and $q$ are edge labels, $a$ is the node label of one of the neighbours of the mother node and $B$ is a node in $D$. The interpretation is that we find an edge labelled $p$ in the host graph which had connected a node $x$ labelled $a$ to the mother node and create a new edge labelled $q$ between $x$ and the node $B$ in the daughter graph. Here flag $d$ works as in Definition \ref{Dncence}. 
\item  $(a,\alpha,b\mid d)$, where $a$ is the node label of one of the nodes in $D$, $b$ is the label of any of the nodes in remaining graph after removing the mother node and $\alpha$ is the label of the new edge connecting $a$ and $b$. The flag $d\in\{0,1,-1\}$ indicates the edge direction. $0$ indicates that the connection is bidirectional, $1$ indicate a directed edge from $b$ to $a$ and  $-1$ indicate a directed edge from $a$ to $b$.
\end{enumerate}
\item []$G_S$ is the start/initial graph,
\item []$R(P)$ is a regular control which specifies the order of application of the productions,
Hence this grammar becomes restricted or non-confluent.
\end{itemize}
\end{definition}
The language represented by this grammar is \[L\text{(}DrncJGG\text{)}={ \{ G \in G_\Delta |G_S \overset{R\text{(}P\text{)}}{\implies}G \} }\] where {$G_\Delta$} is the set of graphs containing only terminal nodes. Hence the language of an $DrncJGG$ graph grammar is a  set of graphs whose nodes have terminal labels, generated by applying a series of productions in the order $p_1, p_2, \cdots, p_n$ where $p_1p_2\cdots p_n$ is a word in the language represented by the regular control $R\text{(}P$).
The following example shows the generation of Wheel graphs using $nc-eNCE-JGG$ graph grammar $ncJW$.
\begin{example}\label{ncJS}
Let $ncJS= \text{(} \Sigma, \Delta, \Gamma, \Omega, P, G_S, R\text{(}P\text{)} $) be an $nc-eNCE$ graph grammar with, $\Sigma = \{ C, c, a\}$, $\Delta = \{ a, c\}$, $\Gamma = \{ \alpha\}$, $\Omega =\{\alpha\}$, $P= \{ p_1, p_2\}$, $G_S$ is a single node with label $C$ and $R\text{(}P\text{)}= p_1^* p_2$.
Figure \ref{pstar} shows the production rules and the associated connection instructions for generating directed star graphs while Figure \ref{dstar} shows the derivation of $S_7$. Since every edge in the graph has the same label ($\alpha$) we have omitted the edge labels in the derivation step. 
\end{example}
\begin{figure}[H]
    \centering
     \begin{tikzpicture}[node distance=1.cm,main node/.style={minimum size = .2cm,circle,fill=black!80 ,draw}, scale=2, every node/.style={scale=1}] 
\node (p1) {$p_1:$}{};
\node[main node] (C1) [right of= p1,label={$C$}]{};
\node (d1) [right of= C1]{$\rightarrow$};
\node[main node] (a) [right  of=d1,label={$a$}]{}; 
\node[main node] (rC) [right  of=a,label={$C$}]{}; 
 \node (c1) [right of=rC,anchor=west]{$C_1=\phi$};
 \draw [thick,stealth-] (a) -- (rC) node [above,midway] {$\alpha$};
 \node (p2)[below of= p1] {$p_2:$}{};
 \node[main node] (C2) [right of= p2,label={$C$}]{};
 \node (d2) [right of= C2]{$\rightarrow$};
\node[main node] (c) [right  of=d2,label={$c$}]{};
\node(d)[right of= c]{};
 \node (c2) [right of=d,anchor=west]{$C_2=\{(c,\alpha,a|-1)\}$};
  \end{tikzpicture}
    \caption{Production rules for directed star graph}
    \label{pstar}
\end{figure}
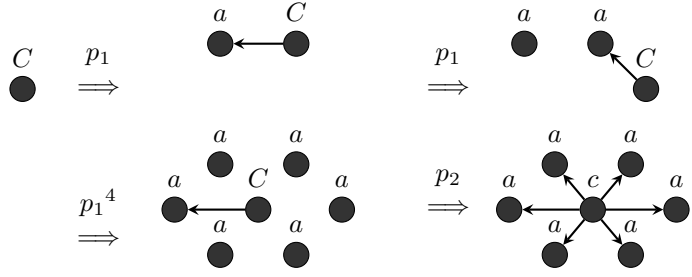
\begin{figure}[H]
    \centering
         \begin{tikzpicture}[main node/.style={minimum size = .2cm,circle,fill=black!80 ,draw}, scale=2, every node/.style={scale=1,minimum height=.2cm,minimum width=1.5cm}] 
     \node[main node] at (1,1)(C1) [label={$C$}]{};
\node at (1.5,1) [label={$p_1$}]{$\Longrightarrow$};    
   \node[main node]at(2.3,1.3)(a) [label={$a$}]{}; 
\node[main node] (C1) at (2.8,1.3) [label={$C$}]{};
\draw [thick,stealth-] (a) -- (C1) ;
\node at (3.8,1) [label={$p_1$}]{$\Longrightarrow$};    
 \node[main node]at (4.3,1.3) (a1) [label={$a$}]{}; 
\node[main node] at (4.8,1.3)(a2)[label={$a$}]{};
\node[main node] at (5.1,1)  (C2) [label={$C$}]{};
\draw [thick,stealth-] (a2) -- (C2) ;
\node at (1.5,0) [label={${p_1}^4$}]{$\Longrightarrow$};
 \node[main node]at (2.3,.5) (a3) [label={$a$}]{}; 
\node[main node] at (2.8,.5)(a4)[label={$a$}]{};
\node[main node] at (3.1,0.2)  (a5) [label={$a$}]{};
\node[main node] at (2.8,-.1)(a6)[label={$a$}]{};
\node[main node] at (2.3,-.1)(a4)[label={$a$}]{};
\node[main node] at (2,.2)(a6)[label={$a$}]{};
\node[main node] at (2.55,.2)(c)[label={$C$}]{};
\draw [thick,stealth-] (a6) -- (c) ;

\node at (3.8,.2) [label={$p_2$}]{$\Longrightarrow$}; 
\node[main node] at (4.5,.5)(a7)[label={$a$}]{};
\node[main node] at (5,.5)(a8)[label={$a$}]{};
\node[main node] at (5.3,0.2)  (a9) [label={$a$}]{};
\node[main node] at (5,-.1)(a10)[label={$a$}]{};
\node[main node] at (4.5,-.1)(a11)[label={$a$}]{};
\node[main node] at (4.2,.2)(a12)[label={$a$}]{};
\node[main node] at (4.75,.2)(c)[label={$c$}]{};
\draw [thick,stealth-] (a7) -- (c) ;
\draw [thick,stealth-] (a8) -- (c) ;
\draw [thick,stealth-] (a9) -- (c) ;
\draw [thick,stealth-] (a10) -- (c);
\draw [thick,stealth-] (a11) -- (c);
\draw [thick,stealth-] (a12) -- (c);
\end{tikzpicture}
\caption{Generation of the star graph $S_7$}
    \label{dstar}
\end{figure}
\section{2D Game Generation}\label{game}
In this section we illustrate how 2D game plots/levels can be generated using graph grammars \cite{2Dgames}. The thrust is given to lock-and-key and platform varieties of games. This is achieved by setting up a puzzle structure, creating a graph grammar to model the layouts and an associated algorithm for arbitrarily producing game levels. The skeleton of the game is a puzzle structure with control points (locks and associated keys) which need to be traversed in a particular order. In order to guarantee that the puzzle is solvable, the designer uses directed edges between the control points and ensures that there is a directed path from the start to the destination.

\par We use the graph grammars introduced in Sections \ref{Dr-ncence} and \ref{djump} to generate game plots. These grammar variants have the extended capability of generating all the components such as locks and their corresponding keys, multiple key based locks, misleading paths, traps, bonus sessions, etc. for setting up a puzzle based graph plot. The plot generation is  demonstrated in Example \ref{puzzle}.  We have also proposed Algorithm \ref{rand_g}  which generates random game plots, given a valid graph grammar of one of these types.  
\begin{example}\label{puzzle}
Consider the  graph grammar $ncG_P= (\Sigma, \Delta, \Gamma, \Omega, P, G_S, R(P) )$, where $\Sigma = \{ S,K,R,G,D,b,e,l,k,m\}$, $\Delta = \{ b,e,l,m,k\}$, $\Gamma = \{ \alpha \}$, $\Omega =\{\alpha\}$, $P = \{ p_1, p_2, p_3, p_4, p_5, p_6, p_7, p_8, p_9 \}$, $G_S$ consists of a single node labelled $S$ and R(P)= $p_1(p_5p_6(p_7p_8^*p_9)^*p_3^*p_4)^*p_2p_3^*p_4$.
Figure \ref{pgame} shows the production rules and the associated connection instructions for generating Puzzle graphs of various forms while Figure \ref{dgame1} and \ref{dgame2} shows the derivation of two such game plots. Since every edge in the graph has the same label ($\alpha$) we have omitted the edge labels in the derivation step.
\end{example}

\begin{figure}[H]
    \centering
    \begin{tikzpicture}
      [node distance=1.5cm,main node/.style={minimum size = .5cm,circle,fill=black!80,anchor=west,draw}, scale=0.7, every node/.style={scale=.65}] 
      \node (p1) {$p_1:$};
      \node[main node] (S) [right  of=p1,label={$S$}]{\textcolor{white}{\small{1}}};
\node (d)[right of= S] {$\rightarrow$};
 \node[main node] (b) [right  of=d,label={$b$}]{\textcolor{white}{\small{2}}};
\node[main node] (G) [right of=b,label={$G$}]{\textcolor{white}{\small{3}}};
\node[main node] (e) [right of=G,label={$e$}]{\textcolor{white}{\small{7}}};

\draw [thick, -stealth] (b.east) -- (G.west)node [above,midway] {$\alpha$};
\draw [thick, -stealth] (G.east) -- (e.west)node [above,midway] {$\alpha$};

\node (p2) [below of=p1,]{$p_2:$};
\node[main node] (G) [right of=p2,label={$G$}]{\textcolor{white}{\small{3}}};
\node (d)[right of= G] {$\rightarrow$};
\node[main node] (K) [right of=d,label={$K$}]{\textcolor{white}{\small{5}}};
\node[main node] (l) [right of=K,label={$l$}]{\textcolor{white}{\small{6}}};
\draw [thick, -stealth] (K.east) -- (l.west)node [above,midway] {$\alpha$};

\node (p3) [below of=p2,]{$p_3:$};
\node[main node] (K) [right of=p3,label={$K$}]{\textcolor{white}{\small{5}}};
\node (d)[right of= K] {$\rightarrow$};
\node[main node] (K) [right of=d,label={$K$}]{\textcolor{white}{\small{5}}};
\node[main node] (l) [right of=K,label={$l$}]{\textcolor{white}{\small{6}}};
\node[main node] (k1) [right of=l,label={$k$}]{\textcolor{white}{\small{7}}};
\draw [thick, -stealth] (K.east) -- (l.west)node [above,midway] {$\alpha$};
  \draw [thick, -stealth] (K.east) -- (l.west)node [above,midway] {$\alpha$};
 \draw [thick] (l.east) -- (k1.west)node [above,midway] {$\alpha$};

\node (p4) [below of=p3,]{$p_4:$};
\node[main node] (K) [right of=p4,label={$K$}]{\textcolor{white}{\small{5}}};
\node (d)[right of= K] {$\rightarrow$};
\node[main node] (k2) [right of=d,label={$k$}]{\textcolor{white}{\small{7}}};
\node (p5) [below of=p4,]{$p_5:$};
\node[main node] (G) [right of=p5,label={$G$}]{\textcolor{white}{\small{3}}};
\node (d)[right of= G] {$\rightarrow$};
\node[main node] (R) [right of=d,label={$R$}]{\textcolor{white}{\small{8}}};
\node[main node] (G) [below of=R,label=below:{$G$}]{\textcolor{white}{\small{3}}};
\draw [thick, -stealth] (R.south) -- (G.north)node [right,midway] {$\alpha$};
\node (d) [below of=p5,]{};
\node (p6) [below of=d,]{$p_6:$};
\node[main node] (R) [right of=p6,label={$R$}]{\textcolor{white}{\small{8}}};
\node (d)[right of= R] {$\rightarrow$};
\node[main node] (K) [right of=d,label={$K$}]{\textcolor{white}{\small{5}}};
\node[main node] (l) [right of=K,label={$m$}]{\textcolor{white}{\small{10}}};
\draw [thick, -stealth] (K.east) -- (l.west)node [above,midway] {$\alpha$};

\node (p7) [below of=p6,]{$p_7:$};
\node[main node] (R) [right of=p7,label={$K$}]{\textcolor{white}{\small{5}}};
\node (d)[right of= R] {$\rightarrow$};
\node[main node] (K) [right of=d,label={$D$}]{\textcolor{white}{\small{9}}};
\node[main node] (l) [right of=K,label={$K$}]{\textcolor{white}{\small{5}}};
\node (p8) [below of=p7,]{$p_8:$};
\node[main node] (K) [right of=p8,label={$D$}]{\textcolor{white}{\small{9}}};
\node (d)[right of= K] {$\rightarrow$};
\node[main node] (K) [right of=d,label={$D$}]{\textcolor{white}{\small{9}}};
\node[main node] (l) [right of=K,label={$l$}]{\textcolor{white}{\small{6}}};
\node[main node] (k) [right of=l,label={$k$}]{\textcolor{white}{\small{7}}};
\draw [thick, -stealth] (K.east) -- (l.west)node [above,midway] {$\alpha$};
  \draw [thick, -stealth] (K.east) -- (l.west)node [above,midway] {$\alpha$};
 \draw [thick, -stealth] (l.east) -- (k.west)node [above,midway] {$\alpha$};
\node (p9) [below of=p8,]{$p_9:$};
\node[main node] (M) [right of=p9,label={$D$}]{\textcolor{white}{\small{9}}};
\node (d)[right of= M] {$\rightarrow$};
\node[main node] (k) [right of=d,label={$k$}]{\textcolor{white}{\small{7}}};
\node(D)[right of=e]{};
\node (C1)[right of=e,anchor=west]{$C_1=\phi$};
\node (D)[below of=e]{};
\node (C2)[right of=D,anchor=west]{$C_2=\{(b,\alpha,\mid 1\mid\alpha,5),(b,\alpha,\mid 1\mid\alpha,6),(e,\alpha,\mid -1\mid\alpha,6),(m,\alpha,\mid 1\mid\alpha,5),(m,\alpha,\mid 1\mid\alpha,6)\}$};
\node (C3)[right of=k1,anchor=west]{$C_3=\{(b,\alpha,\mid 1\mid\alpha,5),(l,\alpha,\mid -1\mid\alpha,7),(b,\alpha,\mid 0\mid\alpha,6),(m,\alpha,\mid -1\mid\alpha,7)\}$};
\node (D)[below of=k1]{};
\node (C4)[right of=D,anchor=west]{$C_4=\{(b,\alpha,\mid 1\mid\alpha,7),(m,\alpha,\mid 1\mid\alpha,7),(l,\alpha,\mid -1\mid\alpha,7)\}$};
\node (D)[below of=D]{};
\node (C5)[right of=D,anchor=west]{$C_5=\{(b,\alpha,\mid 1\mid\alpha,8),(e,\alpha,\mid -1\mid\alpha,3)\}$};
\node (D)[below of=D]{};
\node (D)[below of=D]{};
\node (C6)[right of=D,anchor=west]{$C_6=\{(b,\alpha,\mid 1\mid\alpha,5),(b,\alpha,\mid 1\mid\alpha,10),(G,\alpha,\mid -1\mid\alpha,10)\}$};
\node (D)[below of=D]{};
\node (C7)[right of=D,anchor=west]{$C_7=\{(b,\alpha,\mid 0\mid\alpha,9),(b,\alpha,\mid 1\mid\alpha,5), (l,\alpha,\mid -1\mid\alpha,9)),(l,\alpha,\mid 1\mid\alpha,5)\}$};
\node (D)[below of=D]{};
\node (C8)[right of=D,anchor=west]{$C_8=\{\{(b,\alpha,\mid 1\mid\alpha,9),(l,\alpha,\mid -1\mid\alpha,7),(b,\alpha,\mid 0\mid\alpha,6)\}$};
\node (D)[below of=D]{};
\node (C9)[right of=D,anchor=west]{$C_9=\{(b,\alpha,\mid 0\mid\alpha,6),(l,\alpha,\mid -1\mid\alpha,7)\}$};
  \end{tikzpicture}
    \caption{Production Rules of $ncG_P$}
    \label{pgame}
\end{figure}
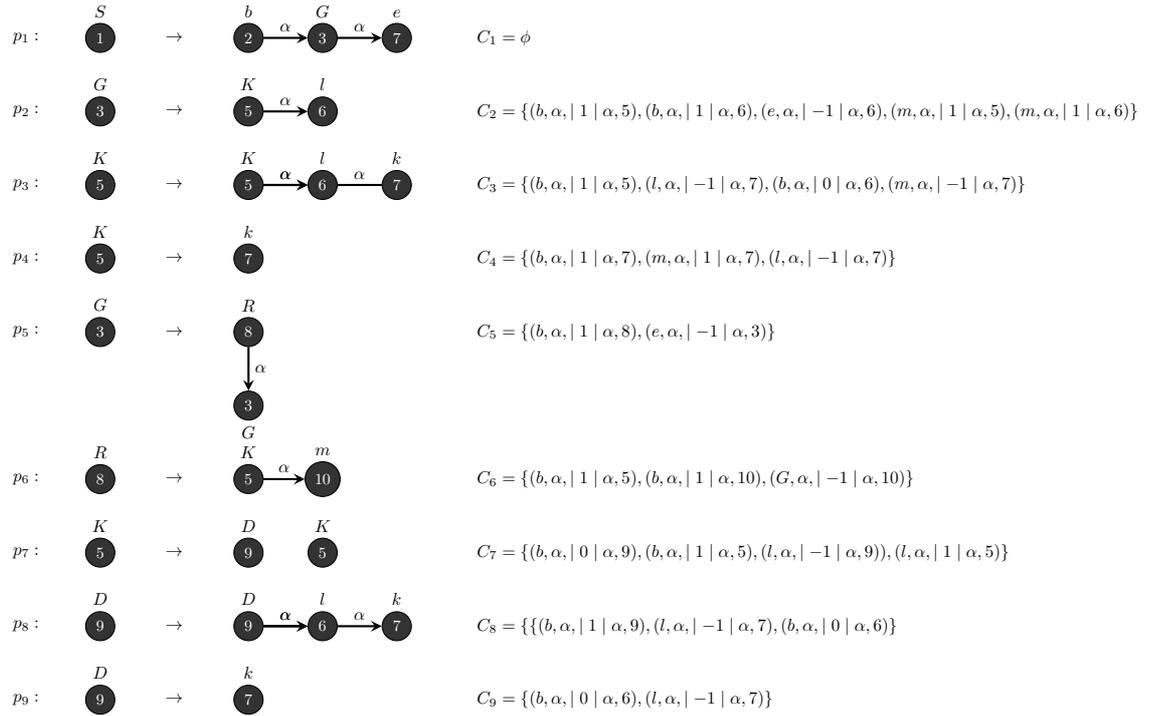
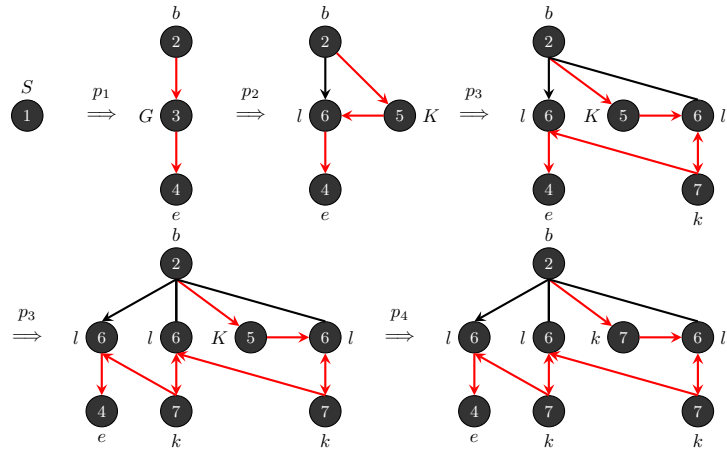
\begin{figure}[H]
    \centering
\begin{tikzpicture}[node distance=1.4cm,main node/.style={minimum size = .5cm,circle,fill=black!80,draw}, scale=1, every node/.style={scale=.7}] 
\node[main node] (1S) [label={$S$}]{\textcolor{white}{\small{1}}};
\node (d1)[right of= 1S,label={$p_1$}] {$\Longrightarrow$};
 \node[main node] (2G) [right  of=d1,label=left:{$G$}]{\textcolor{white}{\small{3}}};
  \node[main node] (2b) [above  of=2G,label={$b$}]{\textcolor{white}{\small{2}}};
  \node[main node] (2e) [below  of=2G,label=below:{$e$}]{\textcolor{white}{\small{4}}};
\draw [thick, color=red, -stealth] (2b.south) -- (2G.north);
\draw [thick, color=red, -stealth] (2G.south) -- (2e.north);

\node (d2)[right of= 2G,label={$p_2$}] {$\Longrightarrow$};
  \node[main node] (3l) [right  of=d2,label=left:{$l$}]{\textcolor{white}{\small{6}}};
  \node[main node] (3K) [right  of=3l,label=right:{$K$}]{\textcolor{white}{\small{5}}};
  \node[main node] (3b) [above  of=3l,label={$b$}]{\textcolor{white}{\small{2}}};
  \node[main node] (3e) [below  of=3l,label=below:{$e$}]{\textcolor{white}{\small{4}}};
\draw [thick, color=red, -stealth] (3b.south east) -- (3K.north west);
\draw [thick, color=red, -stealth] (3K.west) -- (3l.east);
\draw [thick, -stealth] (3b.south) -- (3l.north);
\draw [thick, color=red, -stealth] (3l.south) -- (3e.north);

\node (d3)[right of= 3K,label={$p_3$}] {$\Longrightarrow$};
 \node[main node] (4l1) [right  of=d3,label=left:{$l$}]{\textcolor{white}{\small{6}}};
  \node[main node] (4K) [right  of=4l1,label=left:{$K$}]{\textcolor{white}{\small{5}}};
  \node[main node] (4l2) [right  of=4K,label=right:{$l$}]{\textcolor{white}{\small{6}}};
  \node[main node] (4b) [above  of=4l1,label={$b$}]{\textcolor{white}{\small{2}}};
  \node[main node] (4e) [below  of=4l1,label=below:{$e$}]{\textcolor{white}{\small{4}}};
  \node[main node] (4k) [below  of=4l2,label=below:{$k$}]{\textcolor{white}{\small{7}}};
\draw [thick, color=red, -stealth] (4b.south) -- (4K.north west);
\draw [thick, color=red, stealth-] (4e.north) -- (4l1.south);
\draw [thick, -stealth] (4b.south) -- (4l1.north);
\draw [thick] (4b.south) -- (4l2.north);
\draw [thick, color=red, -stealth] (4K.east) -- (4l2.west);
\draw [thick, color=red, stealth-stealth] (4l2.south) -- (4k.north);
\draw [thick, color=red, stealth-] (4l1.south) -- (4k.north);
\node (D)[below of=1S]{};
\node (D)[below of=D]{};
\node (d4)[below of=D,label={$p_3$}] {$\Longrightarrow$};
 \node[main node] (5l1) [right  of=d4,label=left:{$l$}]{\textcolor{white}{\small{6}}};
  \node[main node] (5l2) [right  of=5l1,label=left:{$l$}]{\textcolor{white}{\small{6}}};
  \node[main node] (5K) [right  of=5l2,label=left:{$K$}]{\textcolor{white}{\small{5}}};
    \node[main node] (5l3) [right  of=5K,label=right:{$l$}]{\textcolor{white}{\small{6}}};
  \node[main node] (5b) [above  of=5l2,label={$b$}]{\textcolor{white}{\small{2}}};
  \node[main node] (5e) [below  of=5l1,label=below:{$e$}]{\textcolor{white}{\small{4}}};
  \node[main node] (5k1) [below  of=5l2,label=below:{$k$}]{\textcolor{white}{\small{7}}};
  \node[main node] (5k2) [below  of=5l3,label=below:{$k$}]{\textcolor{white}{\small{7}}};
\draw [thick, color=red, -stealth] (5b.south) -- (5K.north west);
\draw [thick, color=red, stealth-] (5e.north) -- (5l1.south);
\draw [thick, -stealth] (5b.south) -- (5l1.north);
\draw [thick] (5b.south) -- (5l2.north);
\draw [thick, color=red, stealth-stealth] (5l3.south) -- (5k2.north);
\draw [thick] (5b.south) -- (5l2.north);
\draw [thick, color=red, stealth-stealth] (5l2.south) -- (5k1.north);
\draw [thick, color=red, stealth-] (5l1.south) -- (5k1.north);
\draw [thick, color=red, stealth-] (5l2.south) -- (5k2.north);
\draw [thick, color=red, -stealth] (5K.east) -- (5l3.west);
\draw [thick] (5b.south) -- (5l3.north);

\node (d5)[right of=5l3,label={$p_4$}] {$\Longrightarrow$};
 \node[main node] (6l1) [right  of=d5,label=left:{$l$}]{\textcolor{white}{\small{6}}};
  \node[main node] (6l2) [right  of=6l1,label=left:{$l$}]{\textcolor{white}{\small{6}}};
  \node[main node] (6K) [right  of=6l2,label=left:{$k$}]{\textcolor{white}{\small{7}}};
    \node[main node] (6l3) [right  of=6K,label=right:{$l$}]{\textcolor{white}{\small{6}}};
  \node[main node] (6b) [above  of=6l2,label={$b$}]{\textcolor{white}{\small{2}}};
  \node[main node] (6e) [below  of=6l1,label=below:{$e$}]{\textcolor{white}{\small{4}}};
  \node[main node] (6k1) [below  of=6l2,label=below:{$k$}]{\textcolor{white}{\small{7}}};
  \node[main node] (6k2) [below  of=6l3,label=below:{$k$}]{\textcolor{white}{\small{7}}};
\draw [thick, color=red, -stealth] (6b.south) -- (6K.north west);
\draw [thick, color=red, stealth-] (6e.north) -- (6l1.south);
\draw [thick, -stealth] (6b.south) -- (6l1.north);
\draw [thick] (6b.south) -- (6l2.north);
\draw [thick, color=red, stealth-stealth] (6l3.south) -- (6k2.north);
\draw [thick] (6b.south) -- (6l2.north);
\draw [thick, color=red, stealth-stealth] (6l2.south) -- (6k1.north);
\draw [thick, color=red, stealth-] (6l1.south) -- (6k1.north);
\draw [thick, color=red, stealth-] (6l2.south) -- (6k2.north);
\draw [thick, color=red, -stealth] (6K.east) -- (6l3.west);
\draw [thick] (6b.south) -- (6l3.north);
\end{tikzpicture} 
\caption{Game plot I}
    \label{dgame1}
\end{figure}
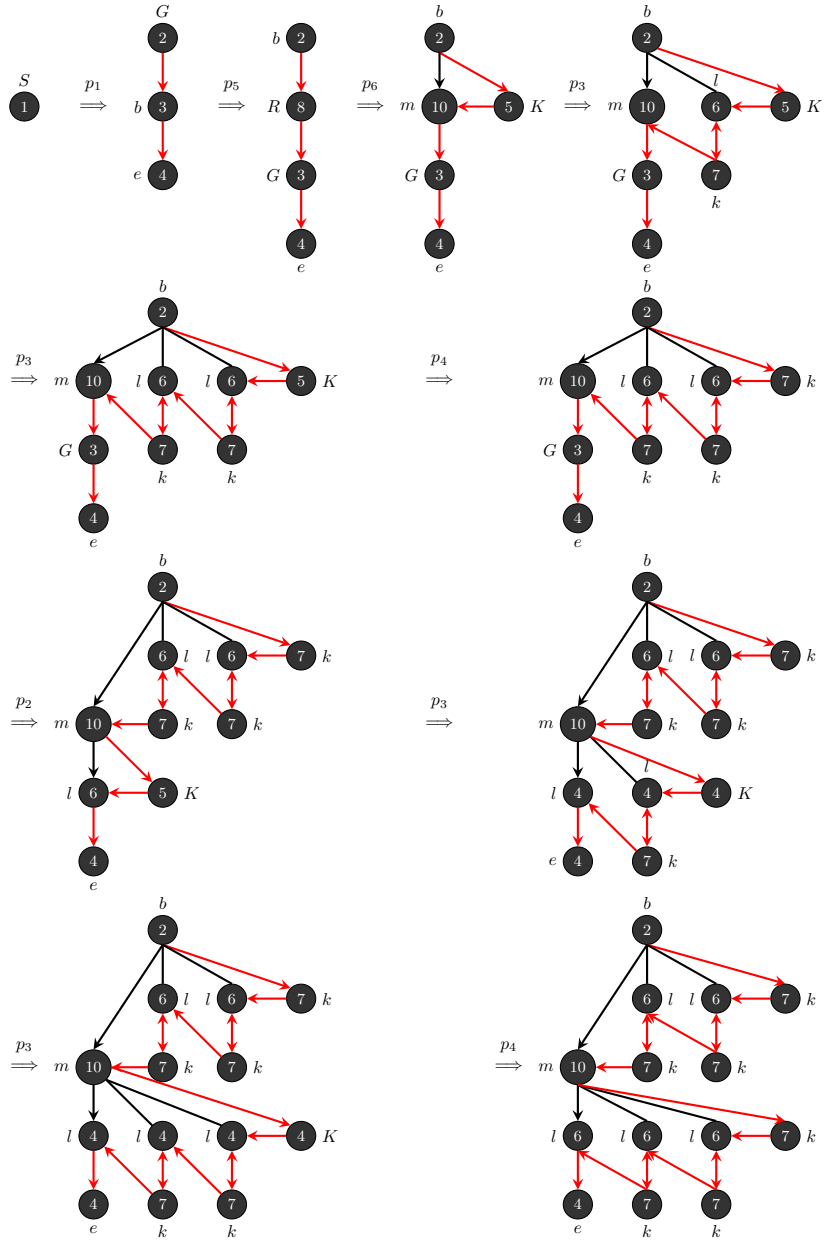
\begin{figure}[H]
    \centering
\begin{tikzpicture}[node distance=1.4cm,main node/.style={minimum size = .5cm,circle,fill=black!80,draw}, scale=0.9, every node/.style={scale=.65}] 
\node[main node] (1S) [label={$S$}]{\textcolor{white}{\small{1}}};
\node (d1)[right of= 1S,label={$p_1$}] {$\Longrightarrow$};
 \node[main node] (2G) [right  of=d1,label=left:{$b$}]{\textcolor{white}{\small{3}}};
  \node[main node] (2b) [above  of=2G,label={$G$}]{\textcolor{white}{\small{2}}};
   \node[main node] (2e) [below  of=2G,label=left:{$e$}]{\textcolor{white}{\small{4}}}; 
\draw [thick, color=red, -stealth] (2b.south) -- (2G.north);
\draw [thick, color=red, -stealth] (2G.south) -- (2e.north);

\node (d2)[right of= 2G,label={$p_5$}] {$\Longrightarrow$};
  \node[main node] (3R) [right  of=d2,label=left:{$R$}]{\textcolor{white}{\small{8}}};
  \node[main node] (3b) [above  of=3R,label=left:{$b$}]{\textcolor{white}{\small{2}}};
    \node[main node] (3G) [below  of=3R,label=left:{$G$}]{\textcolor{white}{\small{3}}};
 \node[main node] (3e) [below  of=3G,label=below:{$e$}]{\textcolor{white}{\small{4}}}; 
\draw [thick, color=red, -stealth] (3b.south) -- (3R.north);
\draw [thick, color=red, -stealth] (3R.south) -- (3G.north);
\draw [thick, color=red, -stealth] (3G.south) -- (3e.north);

\node (d3)[right of= 3R,label={$p_6$}] {$\Longrightarrow$}; 
 \node[main node] (4l) [right  of=d3,label=left:{$m$}]{\textcolor{white}{\small{10}}};
\node[main node] (4K) [right  of=4l,label=right:{$K$}]{\textcolor{white}{\small{5}}};
  \node[main node] (4b) [above  of=4l,label={$b$}]{\textcolor{white}{\small{2}}};
\node[main node] (4G) [below  of=4l,label=left:{$G$}]{\textcolor{white}{\small{3}}};
\node[main node] (4e) [below  of=4G,label=below:{$e$}]{\textcolor{white}{\small{4}}};
\draw [thick, color=red, -stealth] (4b.south) -- (4K.north);
\draw [thick, color=red, -stealth] (4K.west) -- (4l.east);
\draw [thick, -stealth] (4b.south) -- (4l.north);
\draw [thick, color=red, -stealth] (4l.south) -- (4G.north);
\draw [thick, color=red, -stealth] (4G.south) -- (4e.north);

 \node (d4)[right of= 4K,label={$p_3$}] {$\Longrightarrow$};
\node[main node] (5l1) [right  of=d4,label=left:{$m$}]{\textcolor{white}{\small{10}}};
\node[main node] (5l2) [right  of=5l1,label=above:{$l$}]{\textcolor{white}{\small{6}}};
\node[main node] (5K) [right  of=5l2,label=right:{$K$}]{\textcolor{white}{\small{5}}};
 \node[main node] (5b) [above  of=5l1,label={$b$}]{\textcolor{white}{\small{2}}};
\node[main node] (5G) [below  of=5l1,label=left:{$G$}]{\textcolor{white}{\small{3}}};
\node[main node] (5e) [below  of=5G,label=below:{$e$}]{\textcolor{white}{\small{4}}};
\node[main node] (5k) [below  of=5l2,label=below:{$k$}]{\textcolor{white}{\small{7}}};
\draw [thick, color=red, -stealth] (5b.south east) -- (5K.north);
\draw [thick, color=red, stealth-] (5e.north) -- (5G.south);
\draw [thick, -stealth] (5b.south) -- (5l1.north);
\draw [thick] (5b.south) -- (5l2.north);
\draw [thick, color=red, -stealth] (5K.west) -- (5l2.east);
\draw [thick, color=red, stealth-stealth] (5l2.south) -- (5k.north);
\draw [thick, color=red, stealth-] (5l1.south) -- (5k.north);
\draw [thick, color=red, -stealth] (5l1.south) -- (5G.north);
\node (D)[below of=2e]{};
\node[main node] (6b) [below of=D,label={$b$}]{\textcolor{white}{\small{2}}};
\node[main node] (6l2) [below  of=6b,label=left:{$l$}]{\textcolor{white}{\small{6}}};
\node[main node] (6l1) [left  of=6l2,label=left:{$m$}]{\textcolor{white}{\small{10}}};
\node (d5)[left of=6l1,label={$p_3$}] {$\Longrightarrow$};
 \node[main node] (6l3) [right  of=6l2,label=left:{$l$}]{\textcolor{white}{\small{6}}};
 \node[main node] (6K) [right  of=6l3,label=right:{$K$}]{\textcolor{white}{\small{5}}};
 \node[main node] (6G) [below  of=6l1,label=left:{$G$}]{\textcolor{white}{\small{3}}};
 
 \node[main node] (6e) [below  of=6G,label=below:{$e$}]{\textcolor{white}{\small{4}}};
\node[main node] (6k1) [below  of=6l2,label=below:{$k$}]{\textcolor{white}{\small{7}}};
  \node[main node] (6k2) [below  of=6l3,label=below:{$k$}]{\textcolor{white}{\small{7}}};
\draw [thick, color=red, -stealth] (6b.south) -- (6K.north west);
\draw [thick, color=red, stealth-] (6e.north) -- (6G.south);
\draw [thick, color=red, -stealth] (6l1.south) -- (6G.north);
\draw [thick,-stealth] (6b.south) -- (6l1.north);
\draw [thick] (6b.south) -- (6l2.north);
\draw [thick, color=red, -stealth] (6K.west) -- (6l3.east);
\draw [thick, color=red, stealth-stealth] (6l3.south) -- (6k2.north);
\draw [thick] (6b.south) -- (6l3.north);
\draw [thick, color=red, stealth-stealth] (6l2.south) -- (6k1.north);
\draw [thick, color=red, stealth-] (6l2.south east) -- (6k2.north west);
\draw [thick, color=red, stealth-] (6l1.south east) -- (6k1.north west);

 \node (D)[right of=6K]{};
\node (d6)[right of=D,label={$p_4$}] {$\Longrightarrow$};
 \node (D)[right of=d6]{};
\node[main node] (7l1) [right  of=D,label=left:{$m$}]{\textcolor{white}{\small{10}}};
\node[main node] (7l2) [right  of=7l1,label=left:{$l$}]{\textcolor{white}{\small{6}}};
\node[main node] (7l3) [right  of=7l2,label=left:{$l$}]{\textcolor{white}{\small{6}}};
\node[main node] (7k) [right  of=7l3,label=right:{$k$}]{\textcolor{white}{\small{7}}};
\node[main node] (7b) [above of=7l2,label={$b$}]{\textcolor{white}{\small{2}}};
 \node[main node] (7G) [below  of=7l1,label=left:{$G$}]{\textcolor{white}{\small{3}}};
  \node[main node] (7e) [below  of=7G,label=below:{$e$}]{\textcolor{white}{\small{4}}};
\node[main node] (7k1) [below  of=7l2,label=below:{$k$}]{\textcolor{white}{\small{7}}};
  \node[main node] (7k2) [below  of=7l3,label=below:{$k$}]{\textcolor{white}{\small{7}}};

\draw [thick, color=red, -stealth] (7b.south) -- (7k.north west);
\draw [thick, color=red, -stealth] (7l1.south) -- (7G.north);
\draw [thick, color=red, stealth-] (7e.north) -- (7G.south);
\draw [thick,-stealth] (7b.south) -- (7l1.north);
\draw [thick] (7b.south) -- (7l2.north);
\draw [thick, color=red, -stealth] (7k.west) -- (7l3.east);
\draw [thick, color=red, stealth-stealth] (7l3.south) -- (7k2.north);
\draw [thick] (7b.south) -- (7l3.north);
\draw [thick, color=red, stealth-stealth] (7l2.south) -- (7k1.north);
\draw [thick, color=red, stealth-] (7l2.south east) -- (7k2.north west);
\draw [thick, color=red, stealth-] (7l1.south east) -- (7k1.north west);
 \node (D)[right of=6e]{};
 \node[main node] (8b) [below of=D,label={$b$}]{\textcolor{white}{\small{2}}};
\node[main node] (8l2) [below  of=8b,label=right:{$l$}]{\textcolor{white}{\small{6}}};

 \node[main node] (8l3) [right  of=8l2,label=left:{$l$}]{\textcolor{white}{\small{6}}};
 \node[main node] (8k) [right  of=8l3,label=right:{$k$}]{\textcolor{white}{\small{7}}};
  
\node[main node] (8k1) [below  of=8l2,label=right:{$k$}]{\textcolor{white}{\small{7}}};
\node[main node] (8l1) [left  of=8k1,label=left:{$m$}]{\textcolor{white}{\small{10}}};
\node[main node] (8l21) [below  of=8l1,label=left:{$l$}]{\textcolor{white}{\small{6}}};
\node (d7)[left of=8l1,label={$p_2$}] {$\Longrightarrow$};
\node[main node] (8e) [below  of=8l21,label=below:{$e$}]{\textcolor{white}{\small{4}}};
  \node[main node] (8k2) [below  of=8l3,label=right:{$k$}]{\textcolor{white}{\small{7}}};
\node[main node] (8K) [right  of=8l21,label=right:{$K$}]{\textcolor{white}{\small{5}}};
  
\draw [thick, color=red, -stealth] (8b.south) -- (8k.north west);
\draw [thick, color=red, stealth-] (8e.north) -- (8l21.south);
\draw [thick,-stealth] (8b.south) -- (8l1.north);
\draw [thick] (8b.south) -- (8l2.north);
\draw [thick, color=red, -stealth] (8k.west) -- (8l3.east);
\draw [thick, color=red, stealth-stealth] (8l3.south) -- (8k2.north);
\draw [thick] (8b.south) -- (8l3.north);
\draw [thick, color=red, stealth-stealth] (8l2.south) -- (8k1.north);
\draw [thick, color=red, stealth-] (8l2.south east) -- (8k2.north west);
\draw [thick, color=red, stealth-] (8l1.east) -- (8k1.west);
\draw [thick, color=red, -stealth] (8l1.south east) -- (8K.north west);
\draw [thick, color=red, stealth-] (8l21.east) -- (8K.west);
\draw [thick,stealth-] (8l21.north) -- (8l1.south);

 \node (D)[right of=7e]{};
 \node[main node] (9b) [below of=D,label={$b$}]{\textcolor{white}{\small{2}}};
\node[main node] (9l2) [below  of=9b,label=right:{$l$}]{\textcolor{white}{\small{6}}};

 \node[main node] (9l3) [right  of=9l2,label=left:{$l$}]{\textcolor{white}{\small{6}}};
 \node[main node] (9k) [right  of=9l3,label=right:{$k$}]{\textcolor{white}{\small{7}}};
\node[main node] (9k1) [below  of=9l2,label=right:{$k$}]{\textcolor{white}{\small{7}}};
\node[main node] (9l1) [left  of=9k1,label=left:{$m$}]{\textcolor{white}{\small{10}}};
\node[main node] (9l21) [below  of=9l1,label=left:{$l$}]{\textcolor{white}{\small{4}}};
\node (D)[left of=9l1] {};

\node (d8)[left of=D,label={$p_3$}] {$\Longrightarrow$};
\node[main node] (9e) [below  of=9l21,label=left:{$e$}]{\textcolor{white}{\small{4}}};
  \node[main node] (9k2) [below  of=9l3,label=right:{$k$}]{\textcolor{white}{\small{7}}};
\node[main node] (9l22) [right  of=9l21,label={$l$}]{\textcolor{white}{\small{4}}};
 \node[main node] (9k21) [right  of=9e,label=right:{$k$}]{\textcolor{white}{\small{7}}};
\node[main node] (9K) [right  of=9l22,label=right:{$K$}]{\textcolor{white}{\small{4}}}; 
\draw [thick, color=red, -stealth] (9b.south) -- (9k.north west);
\draw [thick, color=red, stealth-] (9e.north) -- (9l21.south);
\draw [thick, stealth-] (9l21.north) -- (9l1.south);
\draw [thick,-stealth] (9b.south) -- (9l1.north);
\draw [thick] (9b.south) -- (9l2.north);
\draw [thick, color=red, -stealth] (9k.west) -- (9l3.east);
\draw [thick, color=red, stealth-stealth] (9l3.south) -- (9k2.north);
\draw [thick] (9b.south) -- (9l3.north);
\draw [thick, color=red, stealth-stealth] (9l2.south) -- (9k1.north);
\draw [thick, color=red, stealth-] (9l2.south east) -- (9k2.north west);
\draw [thick, color=red, stealth-] (9l1.east) -- (9k1.west);
\draw [thick] (9l1.south east) -- (9l22.north west);
\draw [thick, color=red,stealth-stealth] (9l22.south) -- (9k21.north);
\draw [thick, color=red, -stealth] (9l1.south east) -- (9K.north west);
\draw [thick, color=red, stealth-] (9l22.east) -- (9K.west);
\draw [thick, color=red, stealth-] (9l21.south east) -- (9k21.north west);

\node (D)[right of=8e]{};
 \node[main node] (10b) [below of=D,label={$b$}]{\textcolor{white}{\small{2}}};
\node[main node] (10l2) [below  of=10b,label=right:{$l$}]{\textcolor{white}{\small{6}}};

 \node[main node] (10l3) [right  of=10l2,label=left:{$l$}]{\textcolor{white}{\small{6}}};
 \node[main node] (10k) [right  of=10l3,label=right:{$k$}]{\textcolor{white}{\small{7}}};
\node[main node] (10k1) [below  of=10l2,label=right:{$k$}]{\textcolor{white}{\small{7}}};
\node[main node] (10l1) [left  of=10k1,label=left:{$m$}]{\textcolor{white}{\small{10}}};
\node[main node] (10l21) [below  of=10l1,label=left:{$l$}]{\textcolor{white}{\small{4}}};
\node (d8)[left of=10l1,label={$p_3$}] {$\Longrightarrow$};
\node[main node] (10e) [below  of=10l21,label=below:{$e$}]{\textcolor{white}{\small{4}}};
  \node[main node] (10k2) [below  of=10l3,label=right:{$k$}]{\textcolor{white}{\small{7}}};
\node[main node] (10l22) [right  of=10l21,label=left:{$l$}]{\textcolor{white}{\small{4}}};
 \node[main node] (10k21) [right  of=10e,label=below:{$k$}]{\textcolor{white}{\small{7}}};
\node[main node] (10l23) [right  of=10l22,label=left:{$l$}]{\textcolor{white}{\small{4}}};
\node[main node] (10K) [right  of=10l23,label=right:{$K$}]{\textcolor{white}{\small{4}}};
\node[main node] (10k22) [below  of=10l23,label=below:{$k$}]{\textcolor{white}{\small{7}}};

\draw [thick, color=red, -stealth] (10b.south) -- (10k.north west);
\draw [thick, stealth-] (10l21.north) -- (10l1.south);
\draw [thick, color=red, stealth-] (10e.north) -- (10l21.south);
\draw [thick, -stealth] (10b.south) -- (10l1.north);
\draw [thick] (10b.south) -- (10l2.north);
\draw [thick, color=red, -stealth] (10k.west) -- (10l3.east);
\draw [thick, color=red, stealth-stealth] (10l3.south) -- (10k2.north);
\draw [thick] (10b.south) -- (10l3.north);
\draw [thick, color=red, stealth-stealth] (10l2.south) -- (10k1.north);
\draw [thick, color=red, stealth-] (10l2.south east) -- (10k2.north west);
\draw [thick, color=red, stealth-] (10l1.east) -- (10k1.west);
\draw [thick] (10l1.south east) -- (10l22.north west);
\draw [thick, color=red, stealth-stealth] (10l22.south) -- (10k21.north);
\draw [thick] (10l1.south east) -- (10l23.north west);
\draw [thick, color=red, stealth-] (10l21.south east) -- (10k21.north west);
\draw [thick, color=red, -stealth] (10l1.east) -- (10K.north west);
\draw [thick, color=red, stealth-] (10l23.east) -- (10K.west);
\draw [thick, color=red, stealth-stealth] (10l23.south) -- (10k22.north);
\draw [thick, color=red, stealth-] (10l22.south east) -- (10k22.north west);

\node (D)[right of=9e]{};
 \node[main node] (11b) [below of=D,label={$b$}]{\textcolor{white}{\small{2}}};
\node[main node] (11l2) [below  of=11b,label=right:{$l$}]{\textcolor{white}{\small{6}}};

 \node[main node] (11l3) [right  of=11l2,label=left:{$l$}]{\textcolor{white}{\small{6}}};
 \node[main node] (11k) [right  of=11l3,label=right:{$k$}]{\textcolor{white}{\small{7}}};
\node[main node] (11k1) [below  of=11l2,label=right:{$k$}]{\textcolor{white}{\small{7}}};
\node[main node] (11l1) [left  of=11k1,label=left:{$m$}]{\textcolor{white}{\small{10}}};
\node[main node] (11l21) [below  of=11l1,label=left:{$l$}]{\textcolor{white}{\small{6}}};
\node (d8)[left of=11l1,label={$p_4$}] {$\Longrightarrow$};
\node[main node] (11e) [below  of=11l21,label=below:{$e$}]{\textcolor{white}{\small{4}}};
  \node[main node] (11k2) [below  of=11l3,label=right:{$k$}]{\textcolor{white}{\small{7}}};
\node[main node] (11l22) [right  of=11l21,label=left:{$l$}]{\textcolor{white}{\small{6}}};
 \node[main node] (11k21) [right  of=11e,label=below:{$k$}]{\textcolor{white}{\small{7}}};
\node[main node] (11l23) [right  of=11l22,label=left:{$l$}]{\textcolor{white}{\small{6}}};
\node[main node] (11K) [right  of=11l23,label=right:{$k$}]{\textcolor{white}{\small{7}}};
\node[main node] (11k22) [below  of=11l23,label=below:{$k$}]{\textcolor{white}{\small{7}}};

\draw [thick, color=red, -stealth] (11b.south) -- (11k.north);
\draw [thick, color=red, stealth-] (11e.north) -- (11l21.south);
\draw [thick,stealth-] (11l21.north) -- (11l1.south);
\draw [thick,-stealth] (11b.south) -- (11l1.north);
\draw [thick] (11b.south) -- (11l2.north);
\draw [thick, color=red, -stealth] (11k.west) -- (11l3.east);
\draw [thick, color=red, stealth-stealth] (11l3.south) -- (11k2.north);
\draw [thick] (11b.south) -- (11l3.north);
\draw [thick, color=red, stealth-stealth] (11l2.south) -- (11k1.north);
\draw [thick, color=red, stealth-] (11l2.south) -- (11k2.north);
\draw [thick, color=red, stealth-] (11l1.east) -- (11k1.west);
\draw [thick] (11l1.south) -- (11l22.north);
\draw [thick, color=red, stealth-stealth] (11l22.south) -- (11k21.north);
\draw [thick] (11l1.south) -- (11l23.north);
\draw [thick, color=red, stealth-] (11l21.south) -- (11k21.north);
\draw [thick, color=red, -stealth] (11l1.south) -- (11K.north);
\draw [thick, color=red, stealth-] (11l23.east) -- (11K.west);
\draw [thick, color=red, stealth-stealth] (11l23.south) -- (11k22.north);
\draw [thick, color=red, stealth-] (11l22.south) -- (11k22.north);
\end{tikzpicture} 
\caption{ Game plot II with multiple hierarchical level.}
    \label{dgame2}
\end{figure}
\par The graph grammars mentioned in section \ref{Dr-ncence} and \ref{djump} are designed in such a way that it ensures that the game plots generated  correspond to a solvable puzzle graph. This is achieved by designing the  grammar such that after each application of a graph production rule the  connection instructions ensure that the daughter  graph replacing  the mother graph  preserves  the  valid directed path from node $b$ to node $e$. In both the plots, the valid positive path to solve the puzzle/ game level is highlighted with red directed edges that starts from $b$ and ultimately ends in $e$.\\

\RestyleAlgo{ruled}
\SetKwComment{Comment}{/* }{ */}
\begin{algorithm}[H]
\caption{Random Game Plot/Map Generation}\label{rand_g}
\SetAlgoLined
\DontPrintSemicolon
\KwIn{$\{ G_s, P, R(P), limit\}$\Comment*[r]{List of parameters}}    
\KwOut{$ G $ \Comment*[r]{Solvable Random Game Plot/Map}}

    \SetKwFunction{FMain}{GameGen}
    \SetKwFunction{FSub}{Reg}
    \SetKwProg{Fn}{Function}{:}{}
    \Fn{\FMain{$G_s, P, R(P), limit$}\Comment*[r]{Main Function}}
{
        $ RE  \longleftarrow$ Reg$(R(P),limit)$;\\
        Apply $RE$ on $G_s ;$\\
        \textbf{return} the resultant graph $ G; $
}
\textbf{End Function}

\Fn{\FSub{$R(P), limit$}\Comment*[r]{Regular expression string Generation}}
{
        \ForEach{$ * \in R(P) $}
        {
            Replace $*$ with $PRNG(seed)\%limit;$ 
        }
        \textbf{return} $S_P; $ 
}
\textbf{End Function}
\end{algorithm}

\par In order to avoid the tedium of manually generating the game plots we use the  procedural content generation process which is  described  in Algorithm \ref{rand_g}. The function  GameGen() which generates the game plot in the form of a puzzle graph calls the subfunction  Reg() which generates a random string from the given regular expression. The parameters for the GameGen() function are the initial graph $G_s$,  the graph production rule set $P$, the  regular expression $R(P)$ and a natural number, $limit$ which gives a maximum value for  the random numbers to be substituted   in $R(P)$. The parameters $R(P)$ and $limit$ are passed to $Reg()$. In $Reg()$ function a random number is generated for every $'*'$ in $R(P)$ which  eventually returns a random string $S_P$ corresponding to $R(P)$. Here we use a pseudo random number generator which takes a randomly chosen seed value given by the designer. The plots shown in Figure \ref{dgame1} is a result of applying the random string of production rules $p_1p_2p_3^2p_4$ in $G_s$. The random string of production rules corresponding to the plot in Figure \ref{dgame2} is $p_1p_5p_6p_3^2p_4p_2p_3^2p_4$.

\section{Conclusion} \label{con}We have modified  some  of the variants of edge and node controlled embedding graph grammars proposed in \cite{jk_icmicds22,jk_symetry} to facilitate the generation of directed graphs. 
A randomly generated graph plot could serve as the basis for the design of a level of a game. This would help the designer in their effort to develop original and authentic puzzle structures that are solvable. The new approach would enable a game to offer players with a countably infinite number of puzzles, thereby extending the amount of time players would stay playing the game. It would be interesting to actually implement  these game plots to design a game. It also  remains to be seen whether an algorithm can be developed to assess a players skills and accordingly  generate a random game plot with a suitable difficulty level. The graph grammar variant mentioned in section \ref{djump} is specifically meant for generating platform games \cite{adams2014} which involve vertical and horizontal navigation of objects with obstacles in between. Further the possibility of the application of these grammars in creating platform game plots with lock and key elements remains to be investigated.

\section*{Declarations}
\begin{itemize}
\item Funding: No funding was received for conducting this study.
\item Competing interests: The authors has no conflicts of interest to declare that are relevant to the content of this article.
\item Ethics approval: Not applicable
\item Availability of data and materials: Not applicable
\item Code availability: Not applicable
\item Authors' contributions: Both authors contributed to the study conception and design. The first draft of the manuscript was written by Jayakrishna Vijayakumar and both authors commented on previous versions of the manuscript. Both authors read and approved the final manuscript.
\end{itemize}





\bibliography{sn-bibliography}

\end{document}